\documentclass[english]{article}
\usepackage[T1]{fontenc}
\usepackage[latin9]{inputenc}
\usepackage{geometry}
\makeatletter
\date{}

\makeatother

\usepackage{babel}
\begin{document}

\title{A Brief Survey of Non-Residue Based Computational Error Correction}

\author{Sriseshan Srikanth, Bobin Deng, Thomas M. Conte\\
School of Computer Science\\
Georgia Institute of Technology}
\maketitle
\begin{abstract}
The idea of computational error correction has been around for over
half a century. The motivation has largely been to mitigate unreliable
devices, manufacturing defects or harsh environments, primarily as
a mandatory measure to preserve reliability, or more recently, as
a means to lower energy by allowing soft errors to occasionally creep.
While residue codes have shown great promise for this purpose, there
have been several orthogonal non-residue based techniques. In this
article, we provide a high level outline of some of these non-residual
approaches.
\end{abstract}

\section*{Overview}

We first classify various approaches to computational error correction
into two broad categories:
\begin{enumerate}
\item \emph{Temporal Redundancy}. This approach is based on the hypothesis
that the probability of transient errors that occur at the same place
to have temporal multiplicity is very low. In other words, a soft
error occurs infrequently at the same device, and as such, repeated
\emph{measurements} in some manner would serve as an indicator to
the correct computation.
\item \emph{Spatial Redundancy}. This approach is based on the hypothesis
that the probability of multiple identical computations to all be
in error at the same time is very low. In other words, by replicating
a computation, any error in a small fraction of the replicas can be
\emph{masked} / overpowered by the other correct replicas.
\end{enumerate}
These principles, it turns out, are fundamental to any sort of error
correction including computation (ex. arithmetic), storage (ex. memory)
and transmission (ex. networking). Some proposals favor spatial redundancy
over temporal redundancy, some vice versa, and some employ both, depending
upon the target fault model and environment. Given a technique, it
is relatively straightforward to determine presence of temporal and/or
spatial redundancy, as such, we leave this to the interested reader.

Von Neumann \cite{von Neumann} was among the first to propose using
redundant components to overcome the effects of defective devices.
He introduced the now widely used technique of Triple Modular Redundancy
(TMR), which essentially uses three devices instead of one and uses
a majority voter to infer a correct output. To note here is that such
a mechanism can correct a single error (meaning that at least two
of the three devices are not in error), or detect most double errors
(where at least two devices are in error, and their outputs are non-identical).
Even today, variants of such a (single error correct, double error
detect) SECDED error model are in use. 

Also proposed in his work were R-fold modular redundancy (RMR), which
reduces to TMR when R=3; Cascade-TMR, which is essentially a multi-level
TMR (ex. 3 sets of TMR modules and an overall voter) and NAND multiplexing.
The latter replaces each processing unit with multiplexed units containing
N lines for each input and output; the multiplex unit itself has two
stages: executive stage (performs the function of the processing unit,
in parallel) and restorative stage (reduces degradation caused by
the executive stage; acts as a non-linear output amplifier). Years
later, Nikolic et al. \cite{Nikolic} provide a quantitative comparison
of these techniques for fault coverage and area overhead necessary.
In addition, they also evaluate a reconfiguration technique that uses
Configurable Logic Blocks (CLBs) to create fixed function atomic fault
tolerant blocks, and clustering them to achieve global fault tolerance.

While these techniques rank high in terms of robustness, they unfortunately,
require a very significant overhead in terms of area and energy. It
is intuitive that by trading performance, it should be possible to
keep component overhead within reasonable limits to achieve reasonable
reliability. Arithmetic codes are designed to do just that. We borrow
the following classification of arithmetic codes from Wakerly\cite{Wakerly}.
\begin{enumerate}
\item \emph{Separate vs Non-separate.}

\begin{enumerate}
\item \emph{Separate: }The encoding of a datum $X$ is the concatenation
of $X$ and a check symbol computed from $X$ by the function $C$.
The encoding of $X$ is given by $f(X)=<C(X),X>$. Further, upon arithmetic
transformation, the data and check symbols are non-interacting. For
example, $f(X+Y)=<C(X)*C(Y),X+Y>$.
\item \emph{Non-separate:} The encoding of a datum $X$ is simply $C(X)$,
with arithmetic transformation being obtained by a single binary operation
on the encoding. For example, $f(X+Y)=C(X)*C(Y)$.
\end{enumerate}
\item \emph{Systematic vs Non-systematic. }For systematic codes, the encoding
of a datum $X$ has a subfield which equals $X$.
\item \emph{Homomorphic vs Non-homomorphic. }For homomorphic codes, the
data and check symbols are transformed using the same function.
\end{enumerate}
In general, non-separate codes are generally non-systematic (although
exception, i.e., systematic non-separate codes exist \cite{Garner1966}),
and that separate codes are also generally homomorphic (again, exceptions
exist \cite{Wakerly}, however, multiplication is not supported and
issues with implementation exist).

A typical example of a separate code is a residue code. In fact, it
has been observed that all separate codes are equivalent to residue
codes \cite{Peterson1958}. However, we omit details regarding these
as residue codes are not the focus of this article.

The remainder of this article is organized as follows. We first introduce
a class of codes known as AN codes, then move onto summarize a large
class of various parity predicting / circuit implementation dependent
techniques and finally cover other orthogonal approaches to computational
error correction.

\section*{AN Codes}

A typical example of non-separate codes is the AN code \cite{Brown1960}.
The check function constitutes a simple multiplication by a ``check
base'' $A$:

$f(X)=AX;$ $f(X+Y)=AX+AY$

As such, the code is valid under addition and subtraction, with possible
error detection/correction. The crux of such an algorithm relies on
the observation that the sum/difference is also a multiple of $A$.
Error detection is relatively straightforward as it just has to verify
that the sum is divisible by $A$. However, error correction involves
division by $A$ and a series of subsequent subtractions and bit shifts,
the count of which depends upon the number in question; resulting
in a multi-variable-cycle latency for error correction, rendering
it difficult to design efficient computers around. Years later, Liu
\cite{Liu1972} proposed a multi-bit error correction for the AN code,
albeit at a much higher cost.

Failure to support multiplication notwithstanding, AN codes also run
the danger of silent data corruptions due to the inherent possibility
of undetected errors (for example, any erroneous number can pass as
correct if it happens to be a multiple of $A$). For these two reasons,
proposals to augment AN codes have been made. Forin \cite{Forin2014}
introduced static signatures ($B$) to augment the encoding as follows:

$f(X)=AX+B_{X}$, $0<B_{X}<A$ can be arbitrarily chosen for each
$X$.

As such, upon addition of numbers $X$ and $Y$, the sum modulo $A$
should be $B_{X}+B_{Y}$. To detect use of potentially stale (although
correct) registers, a timestamping mechanism was further augmented,
known as ANBD encoding:

$f(X)=AX+B_{X}+D$, where D indicates \emph{version}.

Needless to say, for both ANB and ANBD coded systems, software support
can be leveraged to hasten error detection by static assignments of
signatures and pre-computation of their summations for stack variables.
Further, recent work \cite{SChiffel2010} extends this idea to employ
signatures at the basic block level to verify dynamic control flow
(including non-external function calls) wherever possible, by having
instructions communicate counter values to a watchdog entity.

Yet other approaches suggest efficient encoding for multiplication
and compiler techniques \cite{Wappler2007,Fetzer2009}. However, no
error correction scheme is proposed and fail-stop is the default mode
of operation.

\section*{Micro-architectural / ISA independent Techniques}

A different category of 'codes' exists in that they are circuit dependent.
Their main idea is to \emph{predict }parity transformation with arithmetic
operations and/or careful addition of spatio-temporal redundancy.
This constitutes a relatively large body of work, and we will strive
to provide an intuitive treatment and coverage of such approaches
next.

Sun et al. \cite{Sun2010} combine spatial redundancy and temporal
redundancy to propose an error correcting parallel adder. They realize
a Kogge Stone prefix tree using two Han-Carlson trees, and simply
duplicate the generate/propagate circuitry. For elements off the resulting
critical path, temporal redundancy is utilized. As far as adders are
concerned, their approach is efficient as it is reported to correct
93.76\% of soft errors with an area overhead of 12.23\% and a delay
overhead of 6.41\%. However, no such scheme is presented by them for
multipliers.

We now review the cost of several competing approaches in chronological
order.

Johnson et al. \cite{Johnson1988} examine exploitation of spatial,
temporal and hybrid techniques for the purposes of detection. They
note that duplication with comparison incurs over a 100\% area overhead.
In certain cases, using alternating logic (using compliments) is more
efficient \emph{w.r.t.} hardware cost, however, making a function
self-dual may require a 100\% increase in area. Finally, for the purposes
of temporal redundancy, recomputing with shifted/swapped operands
involve negligible area overhead (extra hardware needed only to compute
input encoding and output comparison). However, the latter two approaches
incur over a 100\% increase in latency. While these early implementation
seem to be high-cost, especially given that no correction is performed,
the ideas are fundamental and are refined over the years.

Hsu et al. \cite{Hsu1992} propose a temporal redundancy technique
that utilizes spatial redundancy in a clever manner. TMR is emulated
but hardware overhead is relatively lowered by using 2 x $\frac{n}{3}$
adders/multipliers and using temporal redundancy when an error is
detected, to achieve correction. Their approach incurs a hardware
overhead of 25\% and a delay penalty of 108\%.

Nikolaidis \cite{Nicolaidis2003} propose efficient parity prediction
techniques to achieve (detection only) low area overhead of 17\% for
carry lookahead adders. As noted by the residue based detection work
of Pan et al. \cite{Pan2008}, Nikolaidis et al. propose \cite{Nicolaidis1994}
using differential logic circuits to implement each cell of array-based
multipliers, and, also propose \cite{Nicolaidis1998} output duplicated
Booth multipliers, again, for detection alone. The latter was improved
upon by Marienfeld et al. \cite{Marienfeld2005} to achieve a hardware
overhead of 35\% for detection in 32 bit multipliers.

Peng et al. \cite{Peng2005} develop a mechanism wherein their adder
stops upon error detection, and using a deadlock detector, reconfigure
the adder. They are able to handle single faults in their 32 bit adders
with an 81\% area overhead, 140\% for 2 faults and 211\% for 3 faults.
Vasudevan et al. \cite{Vasudevan2005} develop error detection in
a carry select adder, with an area overhead of 20\%.

Rao et al. \cite{Rao2006} propose exploiting inherently redundant
computation paths in carry generation blocks of carry lookahead adders
to identify a faulty block. For the generator and propagator circuitry,
time redundancy is used via rotated operands. Ghosh et al. \cite{Ghosh2008}
apply temporal redundancy to the Kogge-Stone adder based on the observation
that even and odd carries are independent. In their two cycle addition
technique, first, one of the correct set of bits (even/odd) are computed
and stored at output register. Second, operands are shifted by one
bit and the remaining sets of bits (odd/even) are computed and stored.

Rao et al. \cite{Rao2008Spatial} further propose spatial redundancy
for specific parallel prefix adders to achieve fault tolerance. Their
design results in area overheads of 85\%, 90\% and 63\% for Brent-Kung,
Kogge-Stone and their hybrid implementations respectively.

Valinataj et al. \cite{Valinataj2007} distribute TMR to protect carries
alone, the premise being that correct carries are sufficient to generate
correct output parities for both addition and multiplication. For
a 32 bit carry lookahead adder and a 32 bit wallace tree multiplier,
their technique requires an area overhead of 115\% and 240\% respectively.
Krekhov et al. \cite{Krekhov2008} propose parity prediction for the
purposes of multi-bit error correction for addition, with roughly
a 100\% area overhead. Mathew et al. \cite{Mathew2010} propose parity
prediction for multi-bit error detection and correction in Galois
Field multipliers with over 100\% of area overhead.

Keren et al. \cite{Keren2008} observe that not all $2^{k}$ outputs
of a $k$-bit output are generally valid outputs for a given combinational
circuit. Instead of using redundancy bits, the input to the checker
is created using the output bits and the input bits, assuming the
function unit is implemented as two independent circuits. Based on
the combinational circuits implemented on an FPGA by them, an average
overhead of 85\% in the number of LUTs was observed.

Dolev et al. \cite{Dolev2013} seek to transform Hamming codes with
arithmetic, by generating codes for the fundamental NAND operation.
The idea is to re-generate the code by performing correction on the
input and then performing NAND and XOR. The XOR could be replaced
with a BCH encoder for multi-bit error correction support.

\section*{Other Orthogonal Techniques}

Banerjee et al. \cite{Banerjee2008} outline an ASIC design flow where-in
the application designer specifies which modules are critical and
which can do with approximate outputs. This is especially relevant
in DSP applications, for example, in FIR filters, certain coefficients
are more important than the others. Post identification via a probabilistic
analysis to determine potential fault locations, series transistors
are added to mitigate potential shorts and parallel transistors are
added to mitigate potential opens.

Blome et al. \cite{Blome2006} propose using a small protected cache
of live register values for better coverage than protected register
files in that protecting the non-state logic (like read/write logic)
of a storage structure is easier for smaller structures. As far as
computation is concerned, time delayed shadow latches are used to
leverage temporal redundancy.

Another class of computational error correction techniques is that
of limiting redundancy to that of detection and using rollback to
an error-free checkpoint to recover. Needless to mention, such a mechanism
comes with its own set of trade-offs and challenges, some of which
are: non-zero error detection latency, recovery latency, degree and
stride of checkpoint placement etc.. We refer the interested reader
to the techniques presented by Habkhi et al. \cite{Tabkhi2008} as
a starting point for a discussion of these issues and related work
in the area. Of further note is the use of checkpoint based recovery
in speculative processors.

\section*{Conclusion}

The need for computational error correction itself has been around
for over half a century, with goals ranging from harsh environments,
manufacturing defects, intermittently reliable devices and/or near
threshold computing. We note several methods to computational error
correction in this article, and observe that these are overall relatively
less efficient (in that they incur more area overhead and/or latency
penalty) when compared with residue codes. To wit, prior work \cite{W=000026H1966,ICRC}
on redundant residue codes achieve computational error correction
for addition, subtraction as well as multiplication in an elegant
manner with a little over 50\% overhead in area and with comparable
performance to their non-redundant equivalent. While this is relatively
superior to the techniques discussed in this article, it is an open
research question as to whether better codes/mechanisms (residue based
or otherwise) exist for computational error correction.

\end{document}